\documentclass[12pt,preprint]{aastex}

\shorttitle{The Magnetic Field Structure of Filament Channels}
\shortauthors{Sheeley, Martin, Panasenco, Warren}

\begin{document}

\title{Using Coronal Cells to Infer the Magnetic Field Structure and Chirality of Filament Channels}

\author{N. R. Sheeley, Jr.\altaffilmark{2}, S. F. Martin\altaffilmark{3},
O. Panasenco\altaffilmark{3}\footnote{Now at Advanced Heliophysics, Pasadena, CA 91106}~,
and H. P. Warren\altaffilmark{2}}

\affil{$^{2}$ Space Science Division, Naval Research Laboratory, Washington DC 20375-5352, USA}

\affil{$^{3}$ Helio Research, La Crescenta, CA 91214, USA}

\begin{abstract}
Coronal cells are visible at temperatures of $\sim$ 1.2 MK in Fe XII coronal images obtained
from the Solar Dynamics Observatory (SDO) and Solar Terrestrial Relations Observatory
(STEREO) spacecraft. We show that near a filament channel, the plumelike tails of these
cells bend horizontally in opposite directions on the two sides of the channel like fibrils
in the chromosphere.  Because the cells are rooted in magnetic flux concentrations
of majority polarity, these observations can be used with photospheric magnetograms
to infer the direction of the horizontal field in filament channels and the chirality
of the associated magnetic field.  This method is similar to the procedure for inferring
the direction of the magnetic field and the chirality of the fibril pattern in filament
channels from H$\alpha$ observations. However, the coronal cell observations are easier to
use and provide clear inferences of the horizontal field direction for heights up to
$\sim$50 Mm into the corona.
    
\end{abstract}


\section{INTRODUCTION}

In a previous paper, \cite{SW_12} described cellular features that were visible in
193\,\AA\ solar images, obtained with the Atmospheric Imaging Assembly (AIA) on the
Solar Dynamics Observatory (SDO) \citep{AIA_11} and in 195\,\AA\ images obtained
with the Extreme Ultraviolet Imagers (EUVI) on the Solar Terrestrial Relations
Observatory (STEREO) \citep{HOW_08}.  Also, using spectrally resolved images
obtained with the Extreme-Ultraviolet Imaging Spectrometer (EIS) on the Hinode
spacecraft \citep{EIS_07,DMW_07}, they found that the cells were most visible in
emission lines of Fe XI-Fe XII whose temperatures are close to 1.2 MK.  At lower
temperatures, the cells were replaced by network structures of the lower atmosphere,
and at higher temperatures, the cells blurred into the more continuous features
of the hot corona.  In addition, the Doppler shifts of these lines showed that the
speed was systematically downward at the boundaries of the cells (by
$\sim$20 km s$^{-1}$) relative to the average speed at the centers of the cells.

The cells occurred in regions of predominantly unipolar magnetic network between
coronal holes and filament channels.  Individual cells had diameters $\sim$30 Mm
that were centered on the flux concentrations, not between them like supergranules
in the photosphere \citep{SL_64}.  As solar rotation carried the cellular regions
across the solar disk, the individual structures were visible as cells only when
they were near central meridian.  At other locations, these features appeared as
tapered plumes projecting $\sim$100 Mm toward the closest limb.  Moreover, when the
cells were located at a solar longitude that was midway between the locations of the
SDO and STEREO-B (or STEREO-A) spacecraft, their plumes appeared to lean toward the
west limb as seen from one spacecraft and the east limb as seen from the other
spacecraft.  Thus, these simultaneous orthogonal views were showing sky-plane projections
of structures that extend radially outward from the Sun.  \cite{SW_12} concluded that
the cells are discrete plumes extending upward from flux concentrations in unipolar
magnetic regions like candles on a cake, and are only visible as bright cells separated
by dark intercellular lanes when observed from above (or at least inward along their
nearly radial magnetic field lines).

In this paper, we show that as one approaches the polarity reversal boundary within
a filament channel, the cellular plumes
bend horizontally along the channel with their tapered ends pointing in opposite
directions on the two sides of the boundary.  Because the plumes are
rooted in flux elements of majority polarity, these oppositely directed features
provide a consistent indication of the magnetic field direction along the channel,
which therefore can be determined from an Fe XII 193\,\AA\ image (or Fe XII 195\,\AA\
image) of the cells and a map of their line-of-sight magnetic fields. 

This method is similar to the one that is often used to infer the
horizontal field direction in H$\alpha$ filament channels from the orientation of
the adjacent chromospheric fibrils and their associated line-of-sight fields
\citep{FOU_71,MBT_94,MARTIN_98,MAR_98,MARLINENG_08}.  However, the Fe XII observations
have several advantages over the H$\alpha$ observations.  First, the coronal plumes
are much larger than the chromospheric fibrils and show the horizontal field structure
without the need for the relatively high spatial resolution required using H$\alpha$
observations.  Second, the coronal observations show the bent plumes at much greater
heights in the atmosphere and provide three-dimensional information when solar rotation
and views from multiple spacecraft (like STEREO-A and STEREO-B) are included.  By
comparison, chromospheric fibrils can only indicate the structure underneath these
plumes.  Third, the Fe XII observations are available on a nearly continuous basis
from the AIA instrument on SDO and the EUVI instruments on STEREO-A and STEREO-B.
\cite{PMVV_12} have recently used AIA 193\,\AA\ observations of coronal cells as an
aid for interpreting the origin of twisting and rotating motions in prominence eruptions,
and \cite{WSS_13} have used the Fe XII images to study the origin of the horizontal
magnetic field in filament channels.  Of course, high-resolution chromospheric
observations in H$\alpha$ 6563\,\AA\ and in Ca II 8542\,\AA\ \citep{RWMW_11} can
be obtained from the ground and used to infer chiralities from the patterns of
chromospheric fibrils and filament threads (barbs) as an alternative to spacecraft
observations.

Finally, we note that \cite{SU_10} used $\it{Hinode}$ X-Ray Telescope (XRT) images
as well as STEREO/EUVI images to study the structure and dynamics of filament channels
prior to the launch of the SDO spacecraft.  Not only did they see cellular plumes
curving along the polarity reversal boundary, but also they noticed that the plume arrangement
was different for east-west filament channels (like the polar crown filament channels)
and for north-south oriented filament channels (like the channels that stretch poleward
from the sunspot belts).  For a sample of six north-south channels observed on twelve
solar rotations, those authors found that the plumes on the east side of the
channel curved toward the polarity reversal boundary, but the plumes on the west side
were straight and oriented nearly perpendicular to the boundary.  More recently,
\cite{SU_12} combined SDO/AIA images and STEREO/EUVI images with a flux-rope model
to study the structure of a polar crown prominence.

In the next section, we describe the magnetic structure and chirality (handedness)
of filament channels by combining AIA 193\,\AA\ coronal
images with virtually simultaneous (averaged over the same 5-minute interval)
photospheric magnetograms, obtained in Fe I 6173\,\AA\ with the Helioseismic and
Magnetic Imager (HMI) on SDO.  In the first three figures, we infer a rotational
magnetic topology from downward views of the filament channel.  Then, in the next
three figures, we look for evidence of this rotational structure in nearly
broadside views of a long lived filament channel.  The last figure of this section
is a time-lapse sequence after the eruption of a filament in this channel.  In the
final section of this paper, we summarize the observations and discuss their
implications.

\section{OBSERVATIONS}

Figure~1 compares an AIA 193\, \AA\ image (top) with an HMI photospheric magnetogram
(bottom) obtained when a filament channel and its adjacent coronal holes were at
the central meridian on 2011 June 17.  Like most of the 193\, \AA\ images in this
paper, this one has been constructed by averaging images over a 5-minute interval
(to increase the signal-to-noise ratio) and then by applying a minimal amount of
sharpening.  The magnetic picture was constructed by averaging magnetograms
obtained over the same 5-minute interval and then applying a minimal amount
of smoothing (to reduce the fine-grain noise and some mixed-polarity field).

In this figure, the filament channel is the elongated
dark region of Fe XII intensity running approximately north-south along the
polarity reversal boundary of the photospheric field.  The channel width is not
precisely defined, but probably extends beyond the dark area to include
cellular plumes that are affected by the axial magnetic field along the
polarity reversal boundary, as evidenced by the curved tails of those tapered
plumes.  By extending the width to include those plumes, we obtain agreement
with the definition used previously for H$\alpha$ filament channels whose widths
include the zones where chromospheric fibrils begin to bend along the polarity
reversal boundary \citep{SMITH_68,MARTIN_98,WAMUG_07}.  We shall use this
definition of Fe XII filament channel in the remainder of our paper.

In Figure~1, coronal cells are visible on the positive and negative sides of the
filament channel, extending from the polarity reversal boundary inside the channel
to the distant coronal holes on each side of the channel.  In this paper, we will focus our
attention on the cellular plumes that lie close to the polarity reversal boundary
and show the influence of the axial magnetic field in the filament channel.
  
On the positive (white)-polarity side of the channel, several
cells are stretched into tapered plumes whose tails bend southward along the
channel.  Because these plumes are rooted in magnetic flux elements of positive
polarity, their tapered tails point along the direction of the magnetic field.
Thus, based on
the directions of these plumes,  we conclude that the field changes from
quasi-radial near the positive-polarity coronal hole to horizontal and pointing
southward along the polarity reversal boundary in the middle of the filament channel.
This corresponds to a leftward rotation of the field and a sinistral chirality,
as typically occurs for such southern-hemisphere filament channels
\citep{MARTIN_98,MAR_98}.
The sense of this rotation continues on the opposite side of the channel where
some fainter cells are visible with their tails pointing northward.  Because these
cells are rooted in elements of negative polarity, their fields are directed southward,
consistent with the southward direction that we inferred from the orientations of
the plumes on the positive-polarity side of the channel.

At this point, we recall the observations of \cite{SU_10} who found that the X-Ray
and XUV structures on the two sides of north-south filament channels were asymmetric
with curved features on the east side of the channel and straight features oriented
nearly perpendicular to the channel on the west side.  At first glance, Figure~1 does
seem to possess such an asymmetry, but with the curved plumes on the west (right)
side and the straighter plumes on the east (left) side, opposite to what \cite{SU_10}
found for their sample of north-south filament channels.

The asymmetry is interesting regardless of whether the curved plumes lie
preferentially on the east side or west side of the channel.  There
is no doubt that the plumes on the two sides of the filament channel are oriented
so that that their tails point in opposite directions along the polarity reversal
boundary, giving a consistent southward direction for the horizontal field along
the boundary.  However, a close inspection of the plumes in Figure~1 reveals that
their components normal to the boundary are pointed in the same direction on the
two sides of the boundary, corresponding to magnetic fields that point toward
the boundary from both sides.  However, if we look closely at the faint plumes
in the dark part of the channel, we can see that these plumes become increasingly
coaligned along the polarity reversal boundary, suggesting that the normal components
of field vanish here.  Thus, Figure~1 supports the idea that there is an asymmetry
across the channel, but that the strength of this asymmetry weakens toward the polarity
reversal boundary and the sign of this asymmetry is opposite to what \cite{SU_10} found
for their selection of north-south filament channels.

Figure~2 shows this southern-hemisphere region 28 days later on 2011 July 15 when it
was again at central meridian.  Differential rotation has caused the filament channel
to be more inclined relative to the north-south direction than it was a rotation
earlier in Figure~1.  Cellular plumes are still visible, oriented with their tails
pointing southward on the positive-polarity side of the channel and northward on
the negative-polarity side.  Because these plumes are rooted in flux concentrations
of majority polarity, their tapered ends point along the field on the positive-polarity
side of the channel and opposite the field on the negative-polarity side.  Again, this
indicates a southward-directed field along the channel and a sinistral chirality.
The normal component of field is still oppositely directed on the two sides
of the channel, but some of the negative-polarity plumes are now oriented at smaller
angles to the polarity reversal boundary than Figure~1 showed 28 days earlier.

Looking closely at the images in Figure~2, we find that we can determine the location
of the polarity reversal boundary more accurately from the Fe XII coronal image than from
the photospheric magnetogram.  In attempting to follow the polarity reversal boundary
through weak-field areas in the magnetogram, we become lost in the fine-scale maze
of mixed polarities.  For this reason, it is often necessary to spatially smooth
photospheric magnetograms in order to determine the location of the larger-scale
polarity reversal boundaries \citep{WRM_11}.  However, in this Fe XII image, the
cellular plumes clearly point the way.  \cite{MARTIN_98,MAR_98} noticed the same effect in
her comparison of photospheric magnetograms and H$\alpha$ images whose chromospheric
fibrils provided a precise definition of the polarity reversal boundary.     

Figure~3 shows a northern-hemisphere region on 2011 September 8.  Here, the tails
of the negative-polarity plumes are stretched southward along the filament channel,
pointing opposite to the direction of the horizontal field.  This orientation corresponds
to a dextral chirality, as expected for such northern-hemisphere regions.  A few
plumes are visible with their tails pointing northward on the positive-polarity
side of the channel.  These orientations are consistent with the dextral chirality
that we inferred from plumes on the negative-polarity side.

Also, we see in Figure~3 that
the tails of the plumes are oriented so that their components along and normal to
the channel have different properties on the two sides of the channel.  Whereas
they are oppositely directed along the channel (corresponding to the same horizontal
direction of the magnetic field), the tails point in the same direction normal to the
channel (corresponding to fields directed away from the polarity reversal boundary on
both sides of the channel).  Recall that the normal components were directed toward
the polarity reversal boundary in Figures~1 and 2.  It is easy to see that this direction,
toward or away from the boundary, depends on the polarity of the bent plumes.  The fields
point toward the boundary if the bent plumes are rooted in positive-polarity field,
and the fields point away from the boundary if the bent plumes are rooted in
negative-polarity field, independent of the chirality of the field.  Also, we
see in Figure~3 that on 2011 September 8, the bent plumes lay on the east side of
the channel, consistent with the result that \cite{SU_10} found.

In this analysis, the magnetogram was necessary to infer the direction of the
horizontal field in the filament channel.  However, note that the Fe XII image
is sufficient to determine the chirality of the filament channel without the need
for a magnetogram.  For example, if we look from south to north along the filament
channel in the Fe XII image in Figure~3, we see the tapered ends of the plumes pointing
toward us on the left side of the channel and away from us on the right side.
If we reverse our direction and look from north to south, the tapered ends of the
plumes still point toward us on our left side and away from us on our right side.
So `toward us on the left' corresponds to a dextral chirality.  For the sinistral
channels in Figures~1 and 2, the plumes have the opposite handedness, pointing
`toward us on the right' regardless of which way we look along the channel.
This sense of chirality corresponds to left-skewed arcades over dextral channels
and right-skewed arcades over sinistral channels \citep{MARTIN_98,MAR_98}.  

The images in Figures~1--3 are typical of SDO and STEREO images obtained in the
past few years during the rising phase of sunspot cycle 24.  The observed
chiralities, dextral in the north and sinistral in the south, satisfy the hemisphere
chirality tendencies that were derived from orientations of chromospheric fibrils
\citep{MBT_94,MAR_98}.  Also, the inference of an axial field along the
channel suggests that the transition from an upward-pointing field
on one side of the channel to a downward-pointing field on the other side is produced
by a rotation of the field as described by \cite{MPEL_08,MBT_94}, rather than by
current-free loops that directly join opposite-polarity fields on the two sides of the
channel.  It is this inferred rotation and its accompanying chiral pattern
that distinguishes the magnetic structure of filament channels from potential field
configurations.  

In most of the Fe XII images that we have examined, the tails of cellular plumes
become tapered and fade out as they approach the polarity reversal boundary
\textbf{from one side of a filament channel}.  An exception is visible in the
lower right corner of Figure~3 where the southward-pointing plumes make a sharp
counterclockwise bend and line up with bright threads that point directly
across the polarity reversal boundary near the right edge of the panel.  These
threads link to a growing active region outside the field of view.  As we shall
see later, eruptive events produce `post flare loops' that extend across the
polarity reversal boundary, and it is possible that these threads in Figure~3
are remnants of intermittent eruptions that occurred in the growing active region.

Next, we look at SDO observations to see if we can find examples of this rotational
structure in broadside views.  Figure~4 shows a three-month old
filament channel in the northern-hemisphere on 2012 April 23.  In this oblique view,
the cellular plumes are inclined in opposite directions, tipping like crossed swords,
toward the east on the north side of the channel where the field is positive
and toward the west on the south side where the field is negative.  Thus, for an
observer standing on the positive-polarity side of the channel, the field would tilt
to the right, corresponding to a dextral chirality.  This tilt seems to change with
distance to the channel, with plumes that are nearly horizontal close to the channel
and more vertical farther away, as we have already seen at lower latitudes.  The clockwise
(right-handed) sense of this rotation is also consistent with a dextral chirality.
Finally, we note that from this perspective, the plumes appear more curved on
the east side of the channel than on the west side, and the normal component of field
points away from the polarity reversal boundary.

Looking eastward and poleward along the filament channel in Figure~4, we see that the
channel bends southward around the trailing end of the negative-polarity region to
form a lower-latitude extension of the channel.  Along the positive-polarity side of
this `switchback', the cellular plumes are directed to the west, corresponding to a
southward directed horizontal field.  Thus, the horizontal field of the filament
channel changed from northward-directed to southward-directed as it passed around
the bend, and preserved the dextral chirality of the field.  It is difficult
to characterize the plumes in this switchback, but if we regard the positive-polarity
plumes to be the curved ones, then the eastward asymmetry would also be preserved
around the bend.
 
Our potential-field source-surface calculations (not shown here) indicate that above
this switchback pattern, the magnetic field has a pseudostreamer geometry, produced
when open fields from the like-polarity coronal holes meet high in the corona
\citep{WSR_07}.  It is similar to the pseudostreamer configuration that occurred
above a switchback configuration of dextral filaments on 2010 August 1, as shown
in Figure~12 of the paper by \cite{PMVV_12}.  Such switchback configurations of
dextral (sinistral) chirality are common in the northern (southern) hemisphere
during the rising phase of the sunspot cycle when large unipolar magnetic regions
trail poleward from their sources in active regions (see \cite{MGY_08} and references
contained therein).

Figure~5 shows the same filament channel 28 days later.  This image is dominated
by positive-polarity plumes whose relatively long tails project outward toward the
limb.  Closer to the polarity reversal boundary, the plumes become progressively
more inclined to the east, corresponding to a dextral chirality and a right-handed
rotation of the field.  On the south side of the channel, the plumes are shorter
and bend sharply to the west, indicating a dextral chirality and
the eastward asymmetry seen 28 days earlier in Figure~4.  Near
the negative-polarity coronal hole, the tails of several plumes extend upward
and are faintly visible in projection across the channel.  Based on their proximity
to the coronal hole, we assume that these plumes point along magnetic field lines
that extend to great heights before returning to the Sun.

This region had a different appearance on 2012 February 27 about a month
after it formed.  As shown in Figure~6, the positive-polarity region contained
many cellular plumes with their tails combed in the northeast direction along
the polarity reversal boundary.  On the south side of the channel, the tails of
the negative-polarity plumes bend southward along the channel, indicating a
dextral chirality and an eastward asymmetry.  Near the negative-polarity
coronal hole, several plumes point directly westward across the channel.  These
loops are fading remnants of the `post-flare loops' that formed after a filament
eruption on 2012 February 24.

This filament channel and its aftermath of successively forming loops
are visible in the time-lapse sequence of AIA 193\,\AA\ images in Figure~7.
Panel (a) shows the filament at 2300 UT on February 23 when a distant segment
had erupted, producing the mound of loops seen under the X-shaped darkening at
the east limb.  Panel (b) shows that by 0700 UT on February 24, the rest of that
long filament had erupted and was replaced by loops that were skewed relative to
the axis of the channel.  Panels (c) and (d) show that as time passed, new loops
formed with less skew and at greater heights over the polarity reversal boundary
(assuming that their heights scale roughly as their lengths).

In effect, the time sequence provides a cut-away view of the field, revealing the
well-known weakening of the axial component with height.  Because the majority
polarity is positive on the west side of the channel (cf. Figure~6), this field has
a dextral chirality.  As shown in Panel (f), only a few loops remained visible
across the channel at 1200 UT on February 25, but many loop legs survived to
produce the `sea of cellular plumes' on the west side of the channel.  Their common
alignment (and a potential-field calculation that is not shown here) suggests that
they are linked to negative-polarity flux farther to the north.

\section{SUMMARY AND DISCUSSION}

In the previous section, we combined AIA Fe XII 193\,\AA\ images with
HMI Fe I 6173\,\AA\ magnetograms to infer the direction of the horizontal magnetic
field in filament channels.  The technique is the same as that used previously for
chromospheric structures; the positive-polarity plumes point in the direction of
the field and the negative-polarity plumes point in the opposite direction.  If the
plumes point to the right when you are looking from the positive-polarity side of
the channel, then the field has a dextral chirality.  If they point to the left,
the chirality is sinistral.  Once this handedness has been determined for one
feature within a chiral system, the chiralities of all the other features within
the system are known.  For a dextral filament channel, the field components of the
overlying arches will be skewed to the left and the threads of the associated
filament barbs will project downward to the right (into the minority-polarity flux
elements of negative-polarity).  For a sinistral channel, these quantities will be
reversed.

We have seen that the cellular plumes within filament channels turn sharply sideways, and
point along the polarity reversal boundary, rather than directly across the boundary to
their nearby counterparts of opposite polarity, as would occur for a potential magnetic
field.  Consequently, these observations are consistent with the general consensus that
the magnetic field of a filament channel is non-potential.

In this paper, we have concentrated on the systematic properties of the component of
field along the filament channel.  However, it is also important to consider the normal
component, which is the starting orientation for plumes before they curve inward toward
the polarity reversal boundary.  As described previously by \cite{SU_10}, the plume
orientations are not symmetric across the channel.  On the two sides of the channel,
the tails of the plumes point in opposite directions along the polarity reversal boundary,
but in the same direction normal to this boundary.  Because these two sets of plumes are
rooted in fields of opposite polarity, the normal component of field must point in opposite
directions on the two sides of the polarity reversal boundary, either toward the
boundary if the bent plumes originate on the positive-polarity side of the channel
or away from the boundary if the bent plumes originate on the negative-polarity side.

These oppositely directed fields raise the question of how the fields on one side
of the channel can join their opposite-polarity counterparts on the other side.
\cite{SU_10} suggested that the fields on the `curved-plume' side of the channel
merge into a flux rope in the channel and that the fields on the `straight-plume'
side go much farther away, either to remote footpoints across the channel or to
their opposite-polarity counterparts elsewhere on the Sun.  Regardless of whether
there is a flux rope in the channel, we have no doubt that the bent plumes indicate
fields that curve sideways to become part of the horizontal field in the channel.
The idea that the fields of the straight plumes extend to great distances is
supported by Figures~6 and 7, which show a `sea of positive-polarity plumes',
whose fields reached over the polarity reversal boundary immediately after
the filament eruption (as indicated by the post flare loops) as well as to
regions farther to the northeast (as suggested by potential field calculations
not shown here.

Vertical views of filament channels like those in Figures~1--3 show that on one side
of the channel the tails of the plumes become narrow and fade out as they bend
asymptotically along the polarity reversal boundary.  It is significant that the
tails fade out while pointing along the boundary rather than directly toward the
boundary.  This implies that the field lines do not cross the polarity reversal
boundary directly in a simple arcade of loops.  If they cross the boundary at all,
they do it $\it{incognito}$ by first merging with the horizontal field on
one side of the channel, and then after passing some distance along
the channel, peeling away from the horizontal field on the other side
of the channel.  {\cite{WSS_13} argued that this
scenario is consistent with the diffusive annihilation of flux at the polarity
reversal boundary, but they did not consider an asymmetry across
the channel.

As mentioned in the previous section, the curved plumes were on the west
side of the filament channel in Figures~1 and 2, rather than the east side as
\cite{SU_10} found.  This counter
example prompted us to search for more examples, and we found several in a brief
survey of SDO/AIA 193\,\AA\ images during June--November 2011.  We found all
combinations --- curved plumes on the east side of the channel, curved plumes on
the west side, and curved plumes on both sides at different locations along the
channel.  We obtained the impression that the specific geometry depends on the
shape of the channel as well as the surrounding distribution of flux on the Sun.
Nevertheless, the asymmetry of the field normal to the channel seemed to persist
with plumes curving toward the polarity reversal boundary on one side of the channel
and straighter plumes pointing away from the boundary on the other side.

The converging plumes may also be telling us that the field lines do
not cross the filament channel at low heights, but instead become linked like barbs
to elements of minority polarity on the same side of the polarity reversal boundary or in
opposite-polarity flux at the end of the filament channel \citep{MBT_94,MARTIN_98,MAR_98}.
\cite{SW_12} found that the amount of minority-polarity flux in the cellular regions
is only $\sim$ 0.1--0.3 times the amount of majority-polarity flux, which suggests that
there may not be enough minority-polarity flux available for the bent plumes.
Nevertheless, it seems likely that the field in a filament channel is mainly horizontal
to heights of $\sim$ 30--60 Mm (or to the top of the filament when it occurs) where
field lines are known to arch across the channel.  We expect to learn more about
these flux connections by looking for changes in the orientations of cellular plumes
and skewed loops during filament eruptions.

We found that cellular plumes are not equally bright on both sides of a filament
channel, and sometimes the plumes are only visible on one side.  Although we do not
know the reason for such intensity variations, several observations suggest that the
cellular intensities are related to the field strengths at the roots of the
cells and the linkage of the cells to regions of opposite magnetic polarity.  First,
enhancements of coronal intensity occur at the boundary of a polar coronal
hole when concentrations of magnetic flux within the hole become linked to new-cycle
active regions, apparently by the reconnection of their fields \citep{SWH_89}.  Second,
during filament eruptions, the cells disappear and are temporarily replaced by transient
coronal holes, as if the field lines that initially linked the cells to regions of
opposite polarity were torn open during the eruptions \citep{SW_12}.  Likewise, as these
open field lines reconnect after the eruption, the cells often reform with an asymmetric
brightness distribution, like that shown in Figure~7.  Third, relatively faint plumes
occur in coronal holes when ephemeral magnetic fields emerge there and become linked
to flux elements of majority polarity in the holes \citep{WS_95,WSD_97,def_97,W_98,R_08}.
However, such transient heatings are unnecessary to see the open-field structures if one
uses a lower-temperature emission line.  SDO/AIA Fe IX/X 171\,\AA\ images show these
relatively cool open-field features as cellular plumes filling in the coronal
hole.\footnote{http://umbra.nascom.nasa.gov/images 2013 May 29}
Thus, the cellular plumes are similar to the classic polar plumes modeled by \cite{NEWHAR_68},
except that the cellular plumes are linked to regions of opposite polarity and are
therefore brighter, hotter, and more closely spaced than polar plumes.

The filament channels in this paper all had chiralities that were consistent with the
`hemisphere chirality tendency', with dextral in the north and sinistral in the south
\citep{LEROY_83,MBT_94,MAR_98,ZIRK_98}.  Thus, their fields made a right-handed (clockwise)
rotation across the northern hemisphere channels and a left-handed (counter-clockwise)
rotation across the southern-hemisphere channels.  When these same channels
are described in terms of magnetic helicity and its associated rotation, the signs are
reversed, with negative helicity and left-handed rotation in the north and positive
helicity and right-handed rotation in the south \citep{DEV_00,WSS_13,BF_84}.  This is
essentially a matter of definition with negative helicity occurring when
the current is directed opposite to the magnetic field (as happens for a dextral
configuration) and with rotation corresponding to the leftward skew of the overlying field
lines.  However, numerous real exceptions to the hemispheric chirality `rule' do exist and
have been studied by many authors \citep{VANB_98,VBPMAC_00,MARWEN_09,YMAC_09,MPBELPS_12,WSS_13}. 
     
It is interesting to consider whether coronal cells (or cellular plumes) can
be regarded as individual domains of force-free magnetic structure that have
spread out from larger and more concentrated sources in active regions.  The Fe XII
images do not show spiral patterns within individual cells like those that
have been calculated previously for sunspots and other discrete flux distributions
\citep{NRBM_71,NR_72,SH_75,CH_77}.  Rather, the images in this paper show
teardrop-shaped tails bending sideways under the influence of the horizontal field.
Thus, it is plausible that the magnetic field of the filament channel consists of
force-free domains, each centered on a magnetic flux concentration and exhibiting
writhe rather than twist.  Where these domains come in contact, their azimuthal
fields would either oppose each other and reconnect, or reinforce each other,
depending on the chirality and polarity of the fields.  Thus, in the northern
hemisphere where the fields are dextral, we would expect horizontal fields
to circulate clockwise (seen from above) around large-scale regions of positive
polarity and counter clockwise around regions of negative polarity.  These
circulations would be reversed in the southern hemisphere.  Similar applications
of Ampere's Law have been proposed recently as an inverse cascade process for
transporting helicity \cite{ANT_09,ANT_12} and as a way of calculating the
evolution of the axial magnetic field in filament channels \citep{WSS_13}.

In this paper, we have concentrated on observations of cellular plumes in
filament channels, and have given little attention to prominences and their eruptions.
However, a popular scenario is that solar magnetic fields in filament channels
involve stresses and helicities that accumulate over time and are released during
filament eruptions.  In the future, we may expect to obtain observational constraints
on such hypotheses by tracking the evolution of coronal plumes in these Fe XII images.

We are grateful to the AIA and HMI science teams for providing observations from the
NASA SDO spacecraft.  We are also grateful to the SECCHI science team for providing EUVI
images of coronal cells and to Nathan Rich (NRL) for his continuing help in the
development of software for observing and analyzing SOHO, STEREO, and SDO images.
Also, we thank the referee for helpful suggestions and for pointing out the
interesting asymmetry across filament channels, as described by \cite{SU_10}.
One of us (NRS) is pleased to acknowledge Y.-M. Wang (NRL) for a variety of help and
advice, including his suggestion that the normal component of field be used in
future numerical studies of filament channels.  NRS is also indebted to C. R. Devore
(NASA/GSFC) and S. K. Antiochos (NASA/GSFC) for useful discussions of force-free fields.
OP is grateful to M. Velli (JPL) for pointing out that coronal cells in a filament
channel follow the pattern of chromospheric fibrils. OP and SFM were supported under NASA grant NNX09AG27G. 
SFM was supported under Predictive Science, Inc. contract NNH 12CF37C. NRS and HPW
acknowledge financial support from NASA and the Office of Naval Research.  
 
\bibliography{ms}
\bibliographystyle{/apj}

\clearpage


\begin{figure}[t]
 \centerline{%
 \includegraphics[clip,scale=0.80]{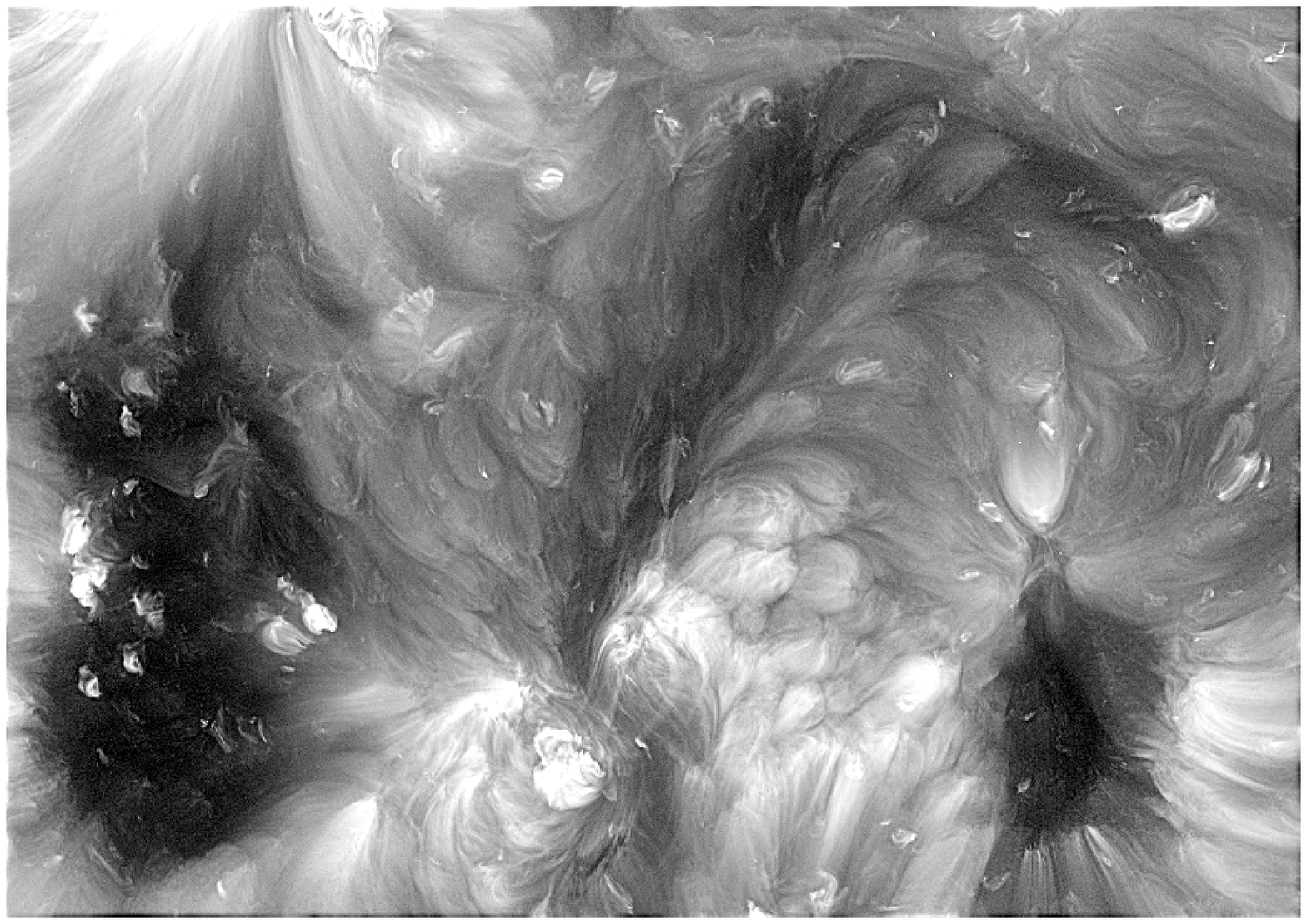}}
 \vspace{0.05in}
 \centerline{%
 \includegraphics[clip,scale=0.80]{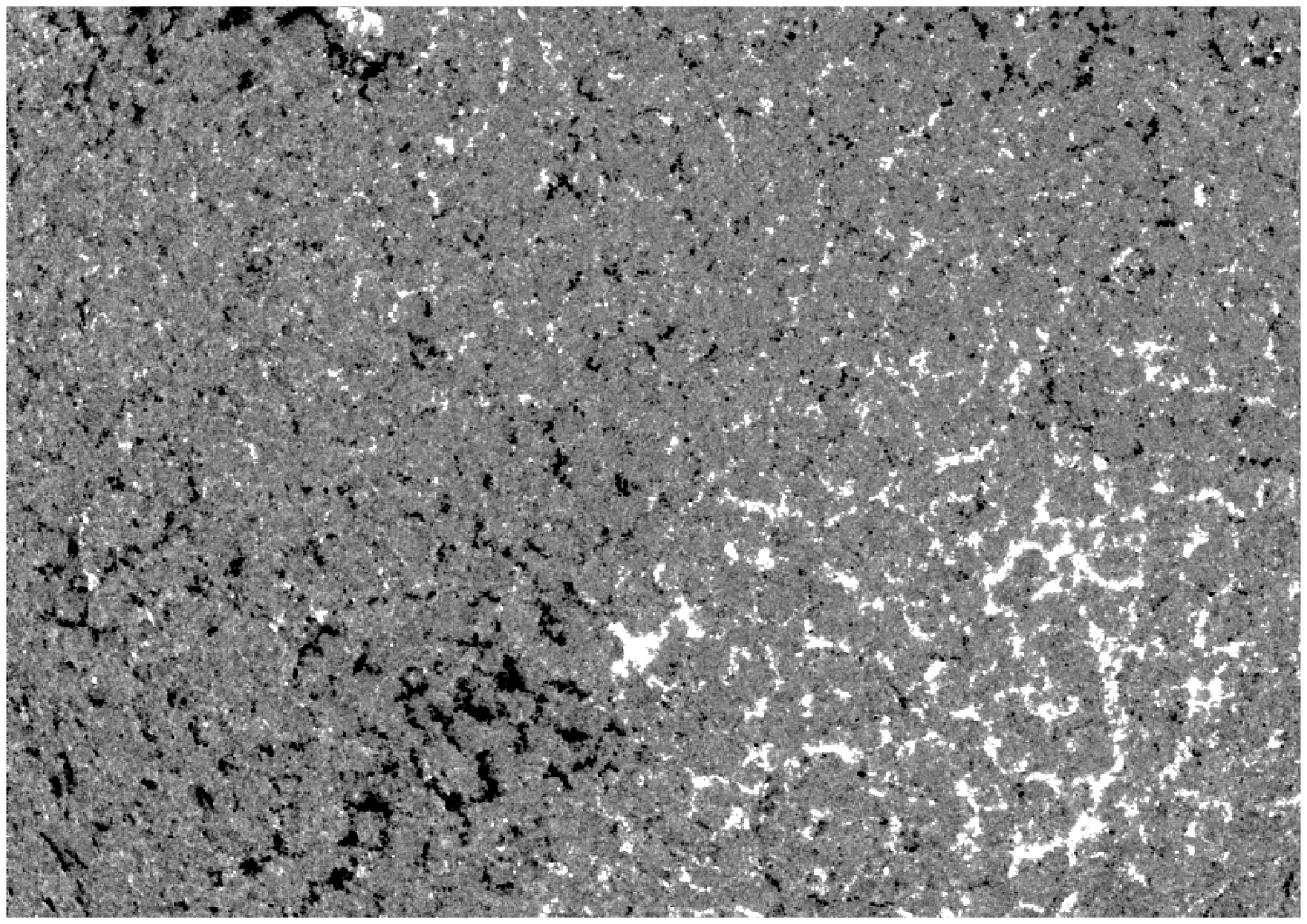}}
 \caption{An AIA 193 \AA\ image and an HMI magnetogram (white positive and black negative),
showing a filament channel between two coronal holes in the southern hemisphere on 2011 June 17.
On each side of the channel, coronal cells are distorted into tadpole shapes with their heads
rooted in elements of majority polarity and their tails pointing in opposite directions
along the channel.  This corresponds to a southward directed field along the channel and a
sinistral chirality.  In all figures of this paper, solar north is up and east is to the left.}
\end{figure}

\begin{figure}[t]
 \centerline{%
 \includegraphics[clip,scale=0.95]{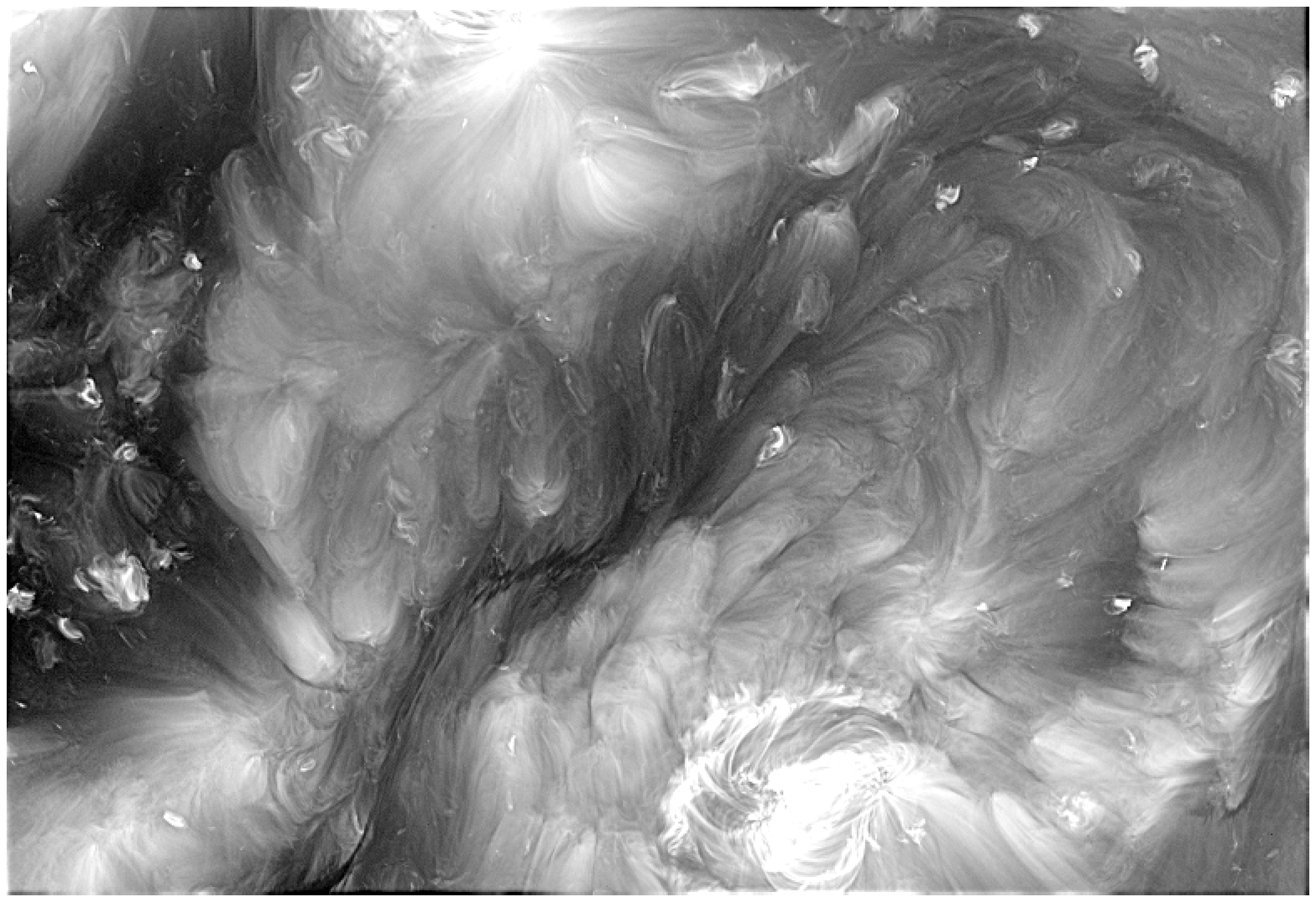}}
 \vspace{0.05in}
 \centerline{%
 \includegraphics[clip,scale=0.95]{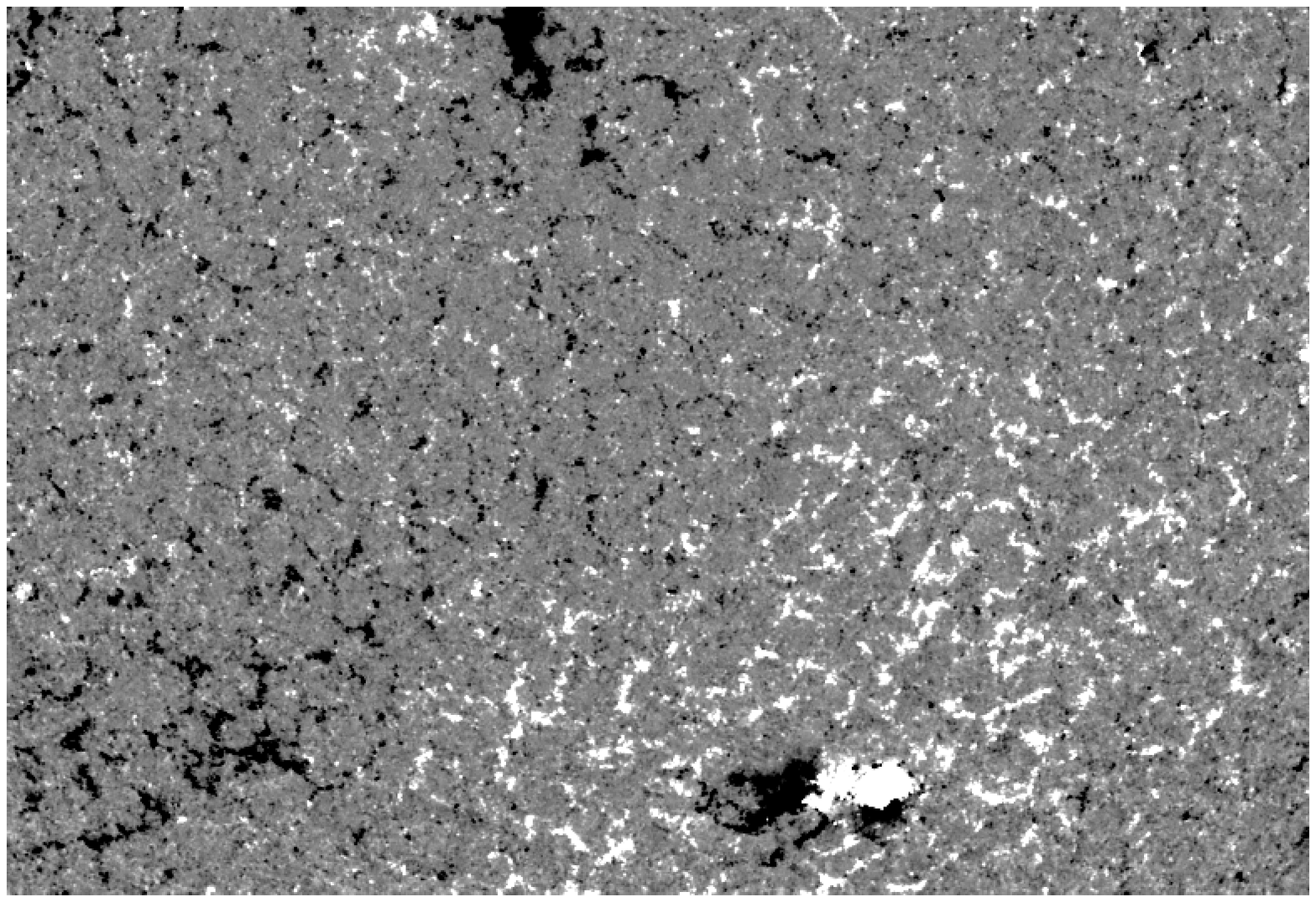}}
 \caption{SDO observations of the region in Fig.~1, obtained 28 days later on 2011 July 15.
Cells are still distorted into tadpoles with their heads rooted in elements of majority
polarity and their tails oriented in opposite directions on the two sides of the channel.
This corresponds to a southward directed field along the channel and a sinistral chirality.}
\end{figure}

\begin{figure}[t]
 \centerline{%
 \includegraphics[clip,scale=0.90]{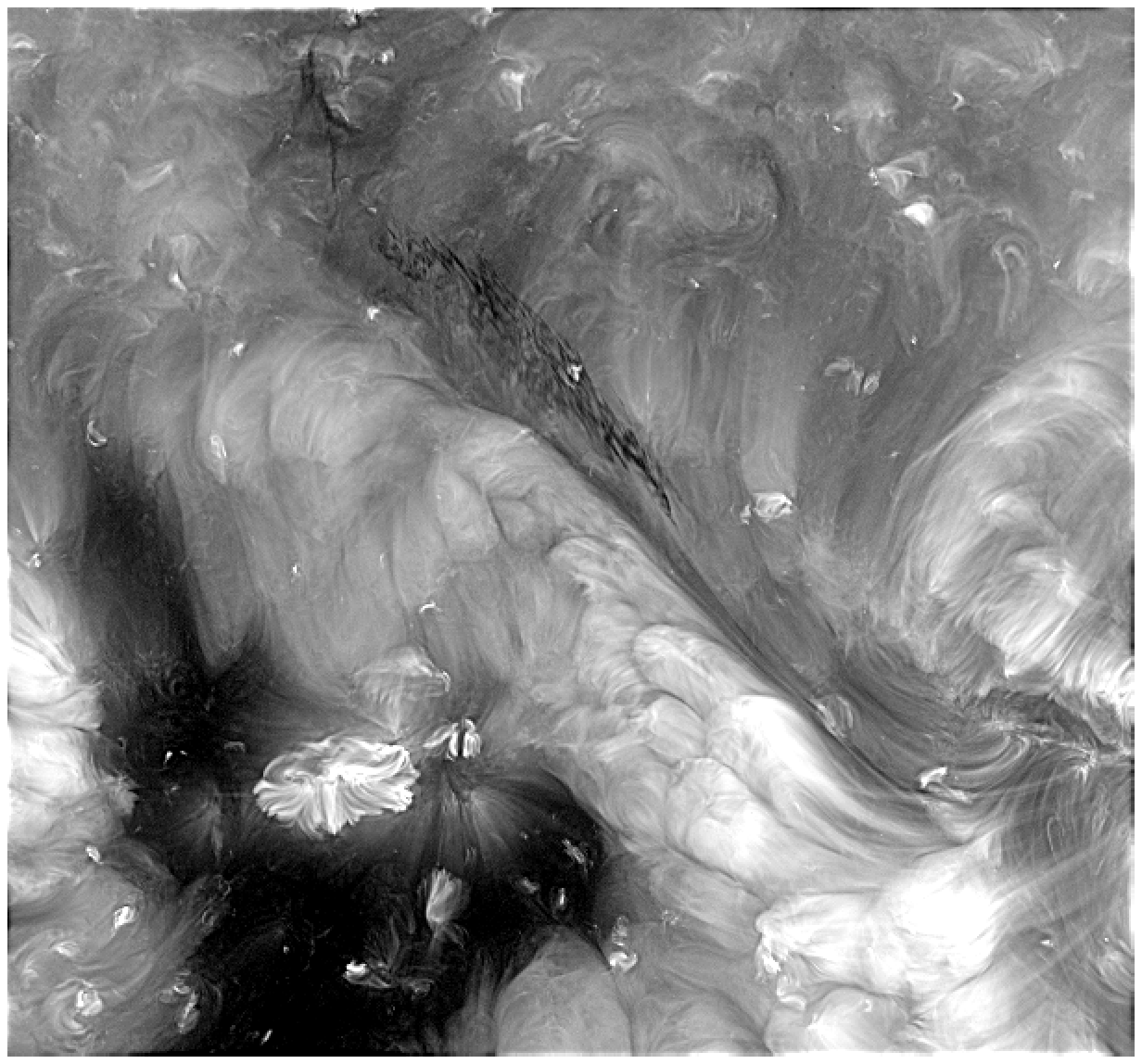}}
 \vspace{0.05in}
 \centerline{%
 \includegraphics[clip,scale=0.90]{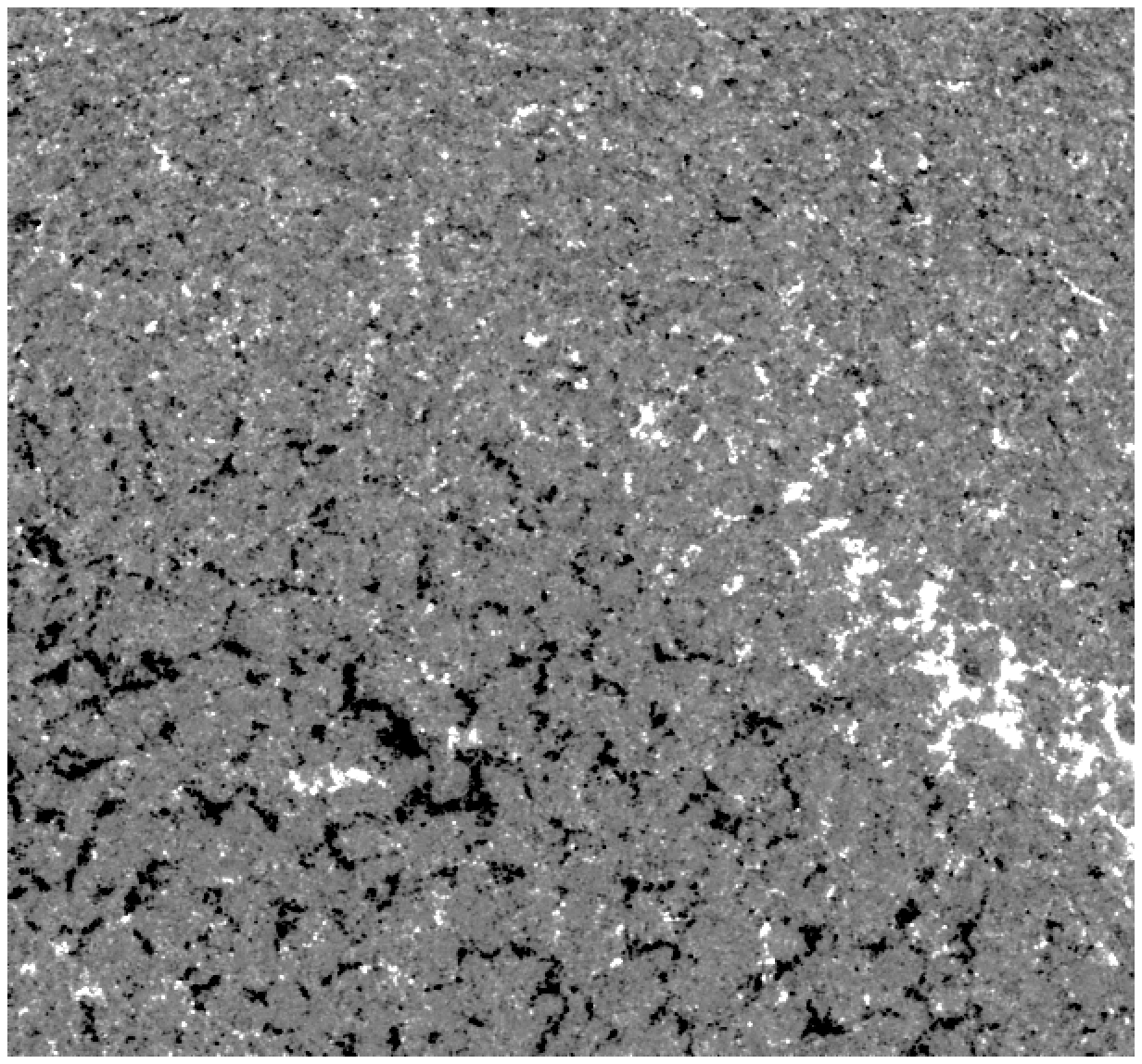}}
 \caption{SDO images of a northern-hemisphere region on 2011 September 08.
Cells, rooted in elements of negative polarity, are distorted into tadpole shapes with
their tails pointing southward along the channel.  This corresponds to a northward
field along the channel and a dextral chirality.}
\end{figure}

\begin{figure}[t]
 \centerline{%
 \includegraphics[clip,scale=0.85]{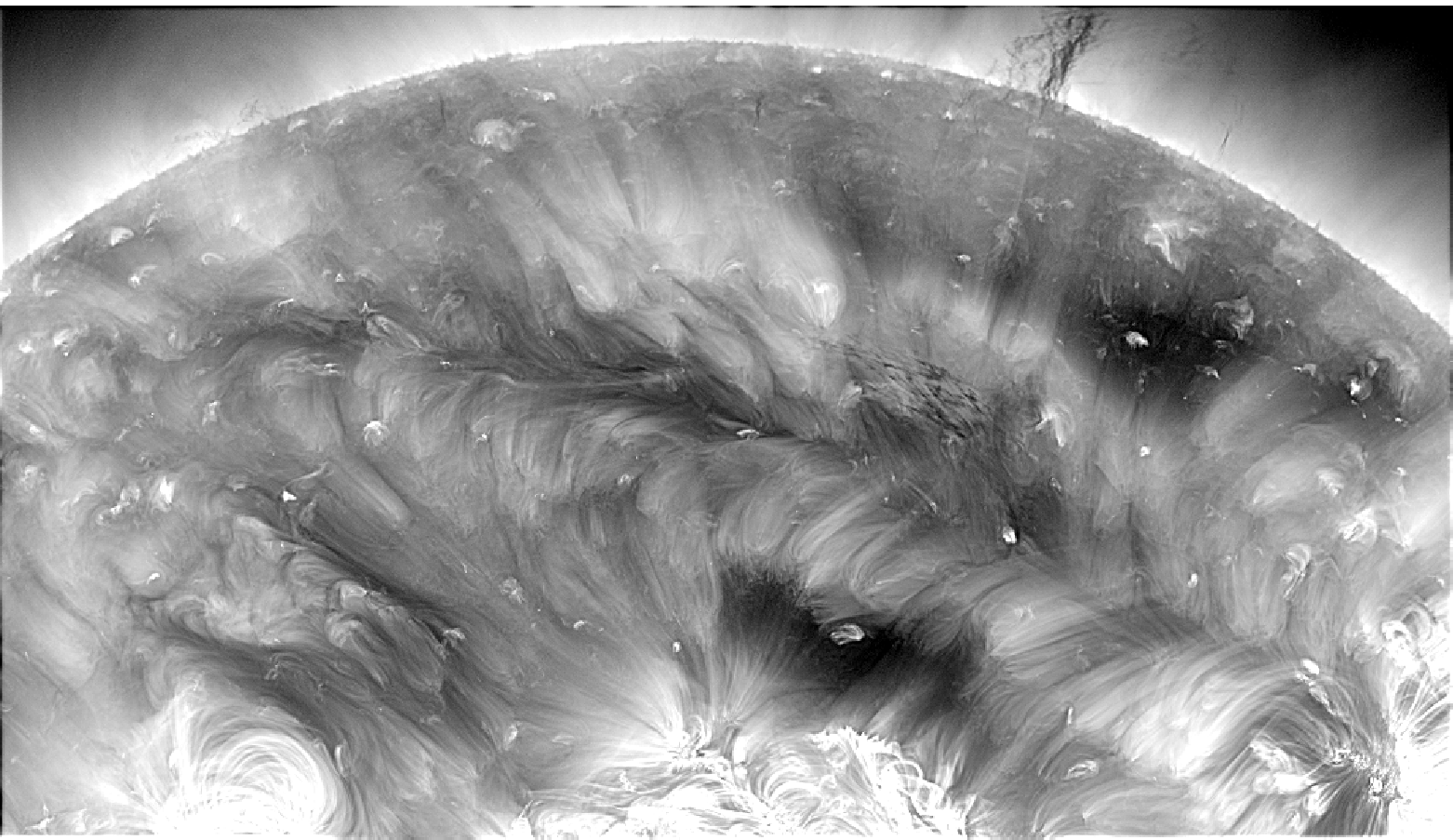}}
 \vspace{0.05in}
 \centerline{%
 \includegraphics[clip,scale=0.85]{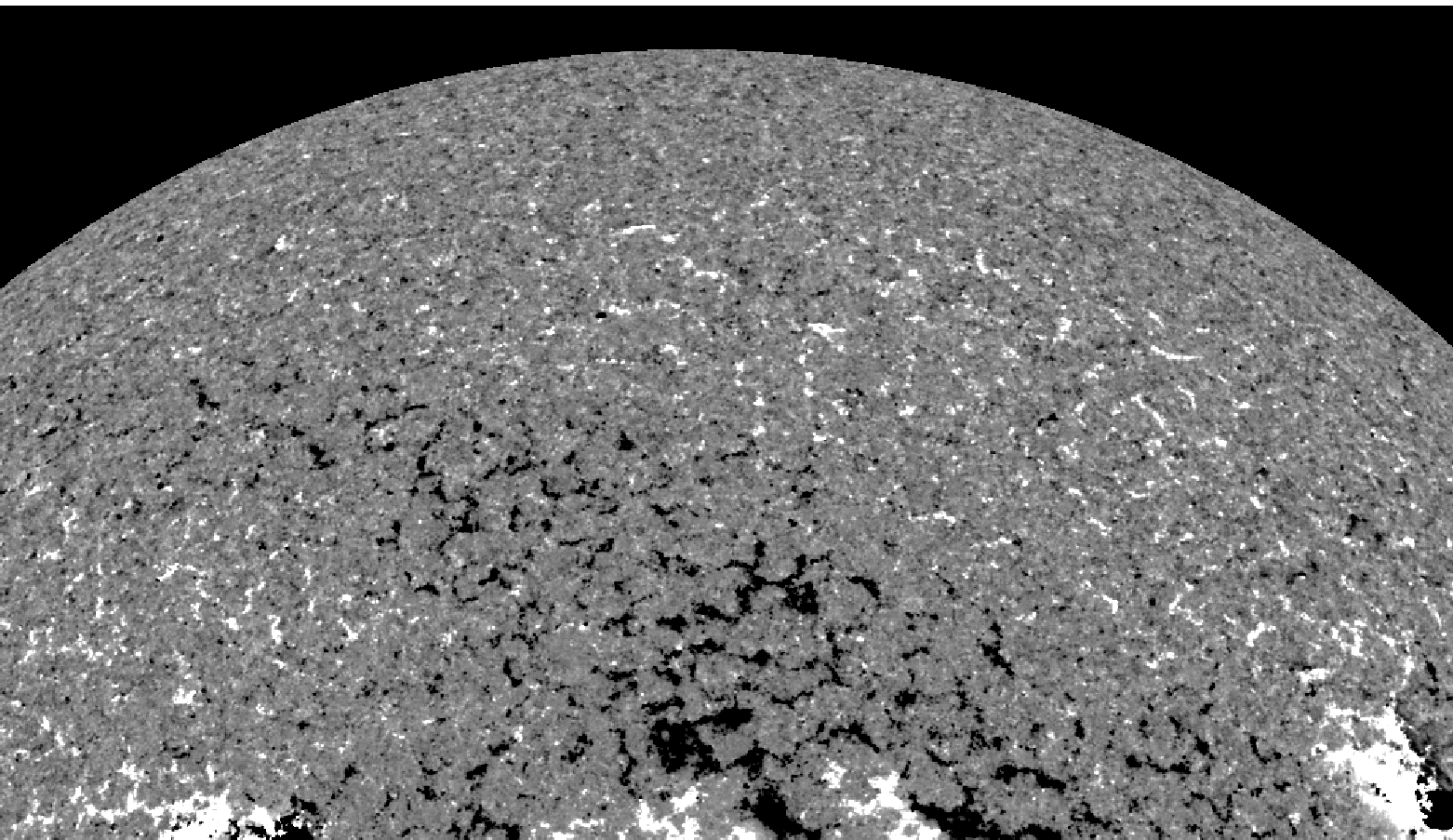}}
 \caption{SDO images of a northern-hemisphere filament channel on 2012 April 23, showing
cellular plumes leaning in opposite directions on the two sides of the channel, corresponding
to a dextral chirality.  This chirality is preserved around the switchback at the trailing
end of the channel where positive-polarity plumes at lower latitude point to the west.}
\end{figure}

\begin{figure}[t]
 \centerline{%
 \includegraphics[clip,scale=0.85]{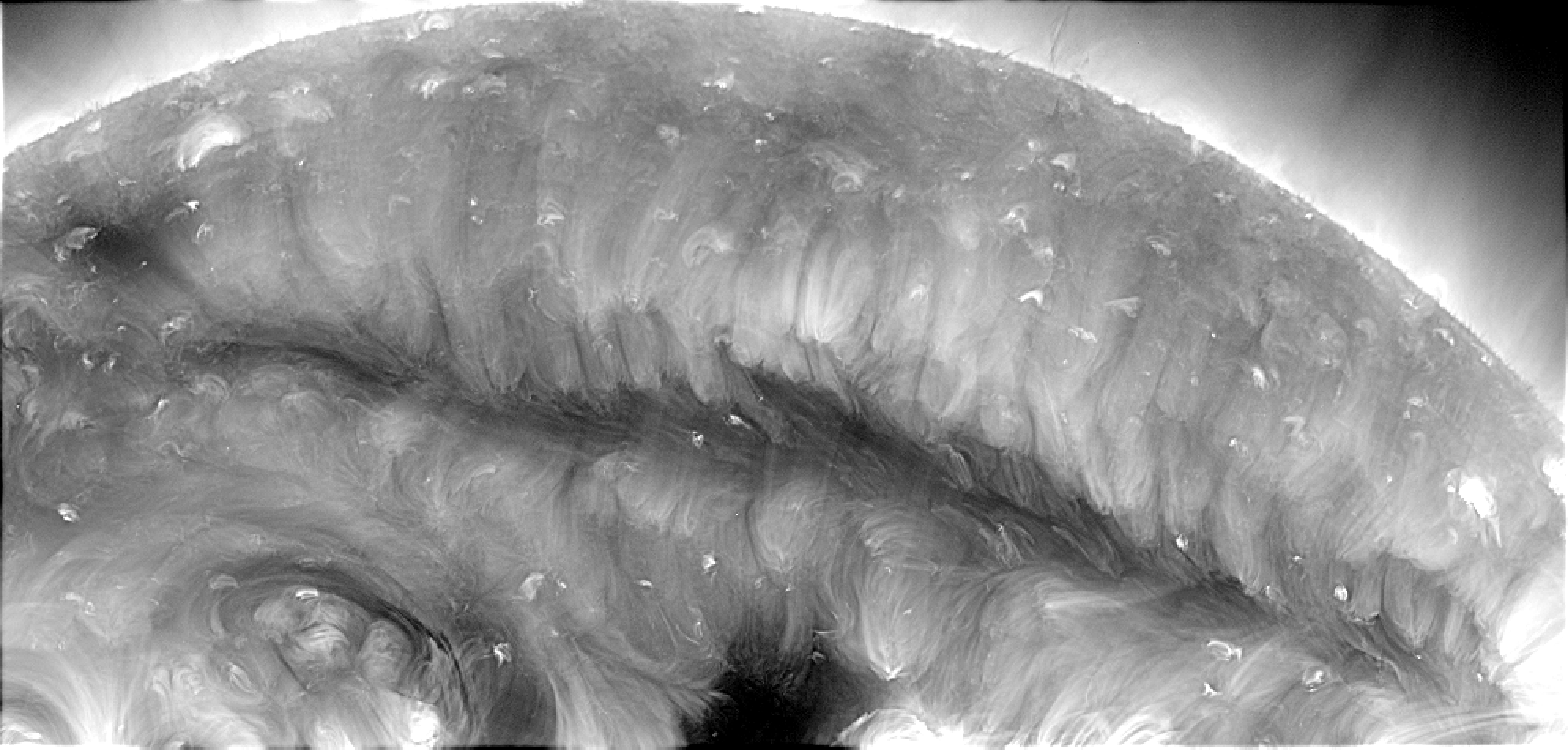}}
 \vspace{0.05in}
 \centerline{%
 \includegraphics[clip,scale=0.85]{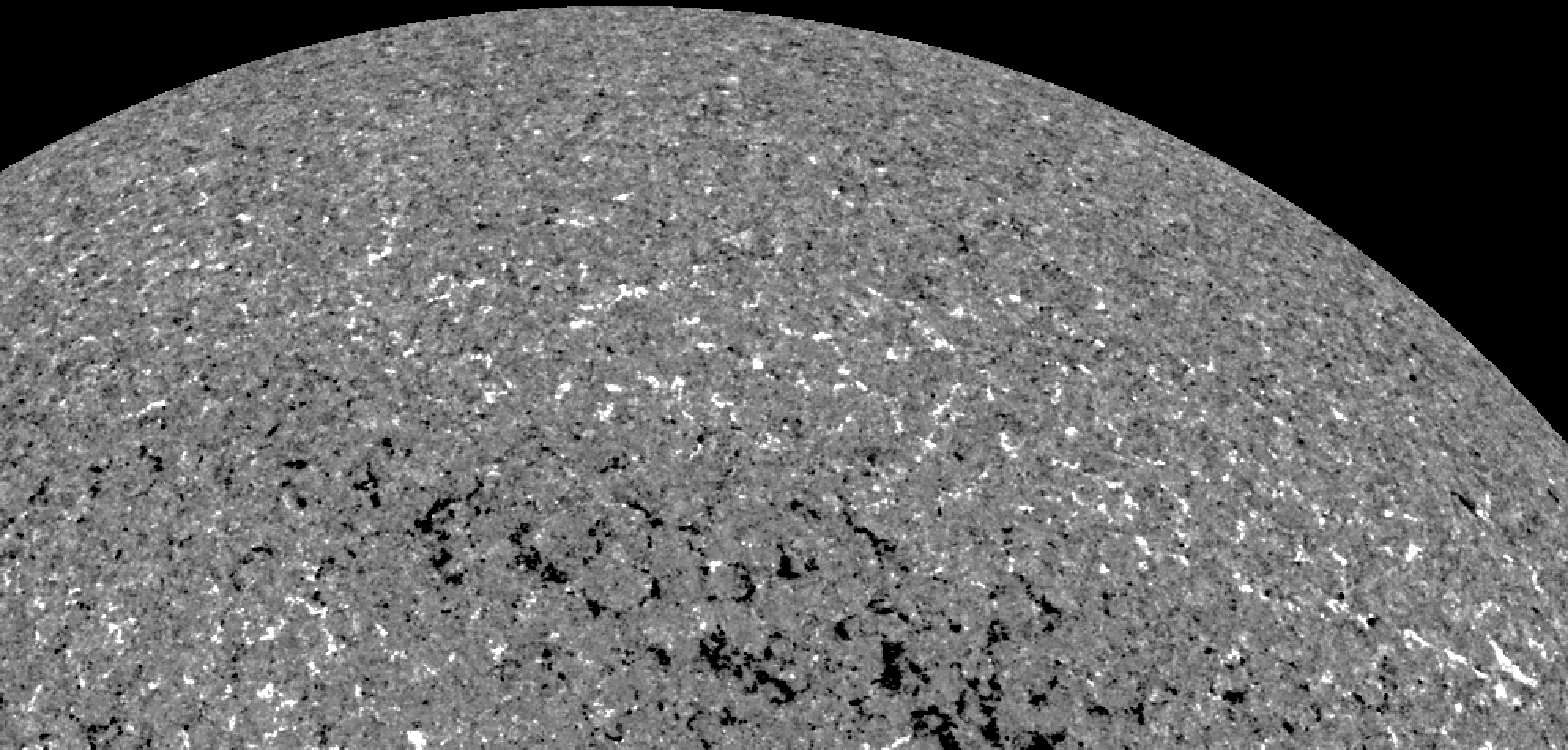}}
 \caption{The region in Fig.~4, 28 days later on 2012 May 21 when the plumes on
the positive-polarity side of the channel appear to be lined up vertically like a row of teeth.
A close examination reveals that these positive-polarity plumes occur in several rows with
fainter, more tilted plumes in the foreground next to the channel and brighter, more
vertical plumes in the back - suggesting a rotation of the field and a dextral chirality.}
\end{figure}

\begin{figure}[t]
 \centerline{%
 \includegraphics[clip,scale=0.80]{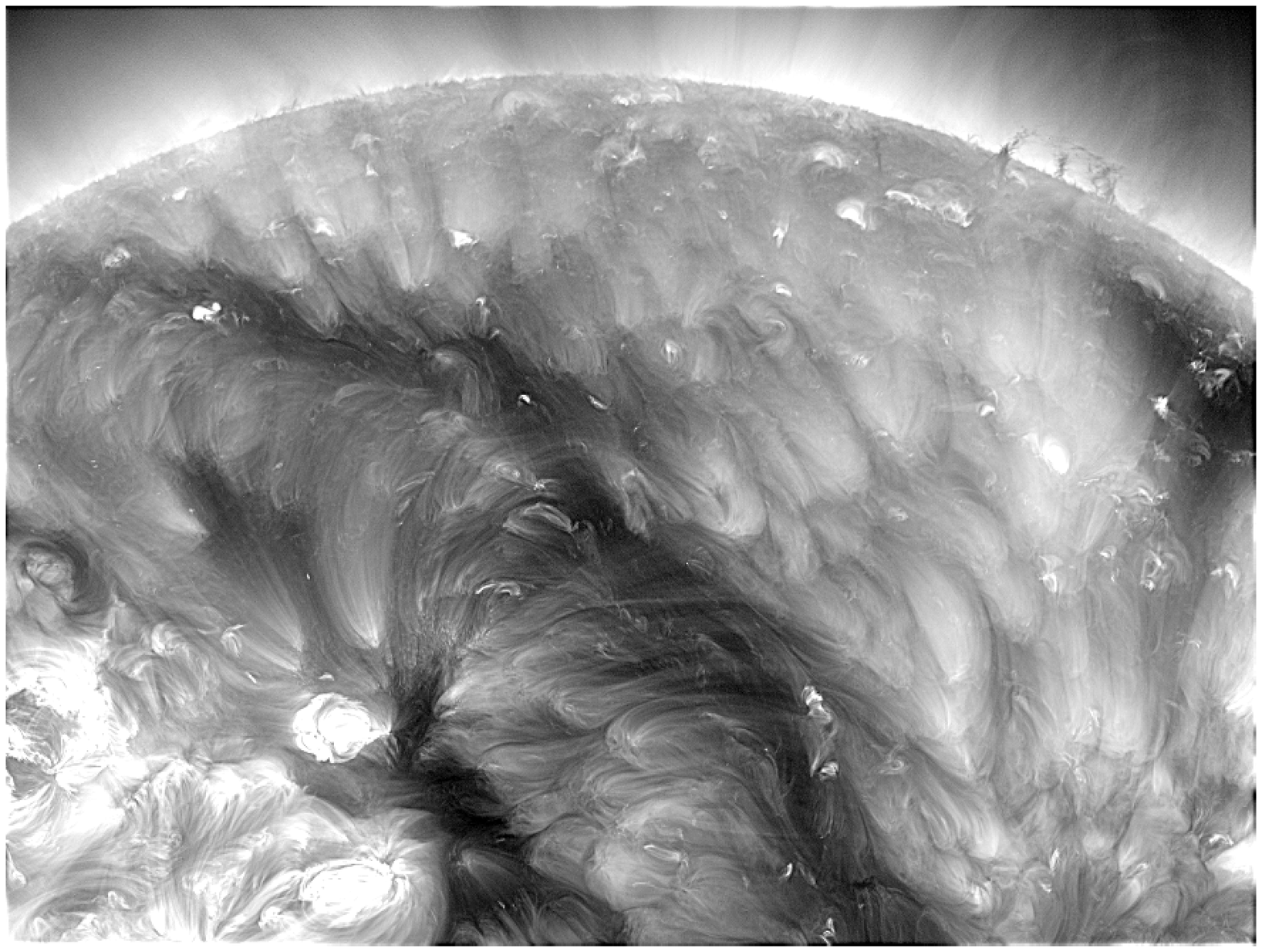}}
 \vspace{0.05in}
 \centerline{%
 \includegraphics[clip,scale=0.80]{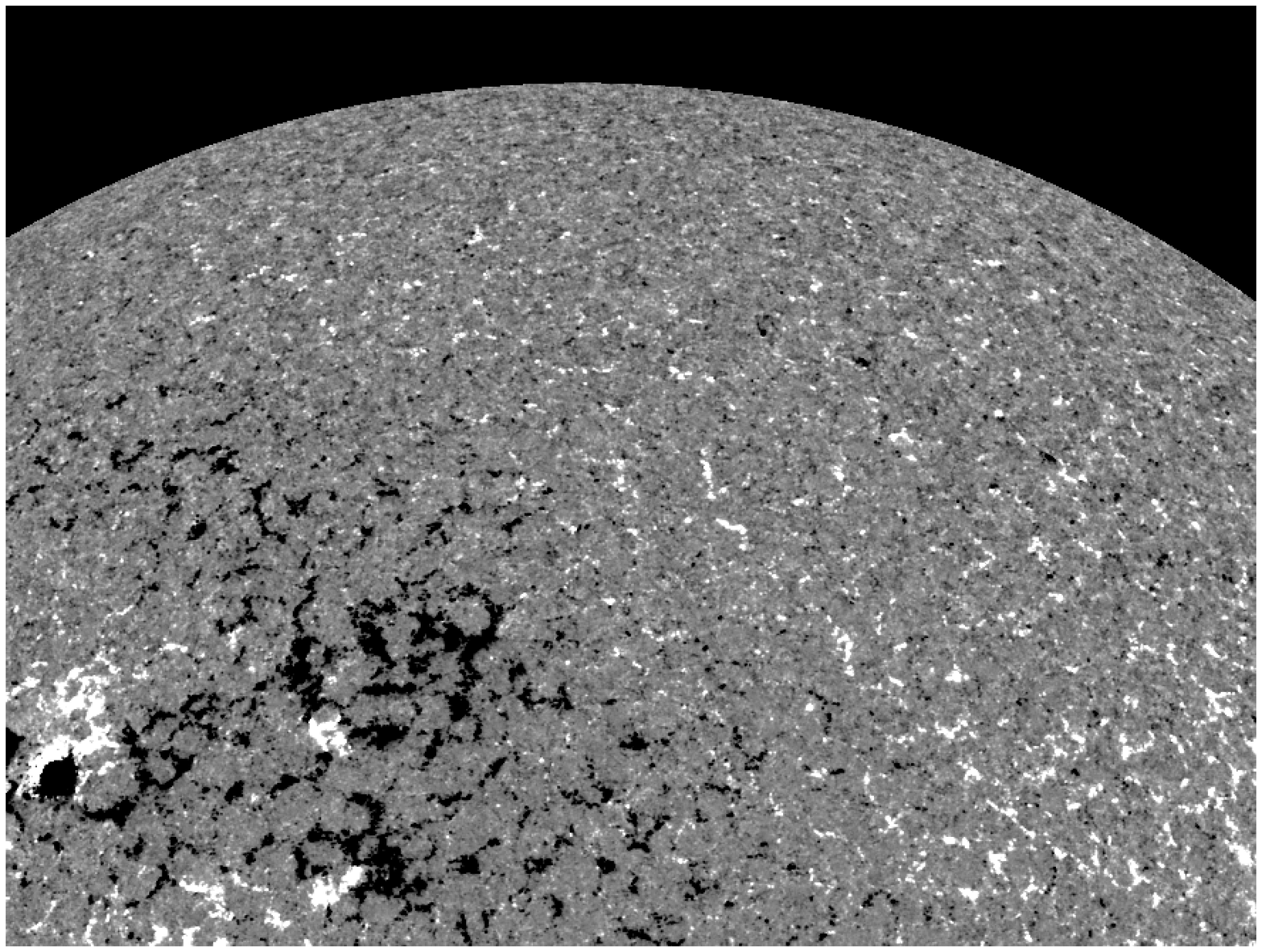}}
 \caption{The filament channel from Figs. 4 and 5, but seen on 2012 February 27, when many positive-polarity
plumes pointed northward parallel to the channel and the tails of some plumes near
the negative-polarity coronal hole stretched westward across the channel.}
\end{figure}

\begin{figure}[t]
 \centerline{%
 \includegraphics[clip,scale=0.95]{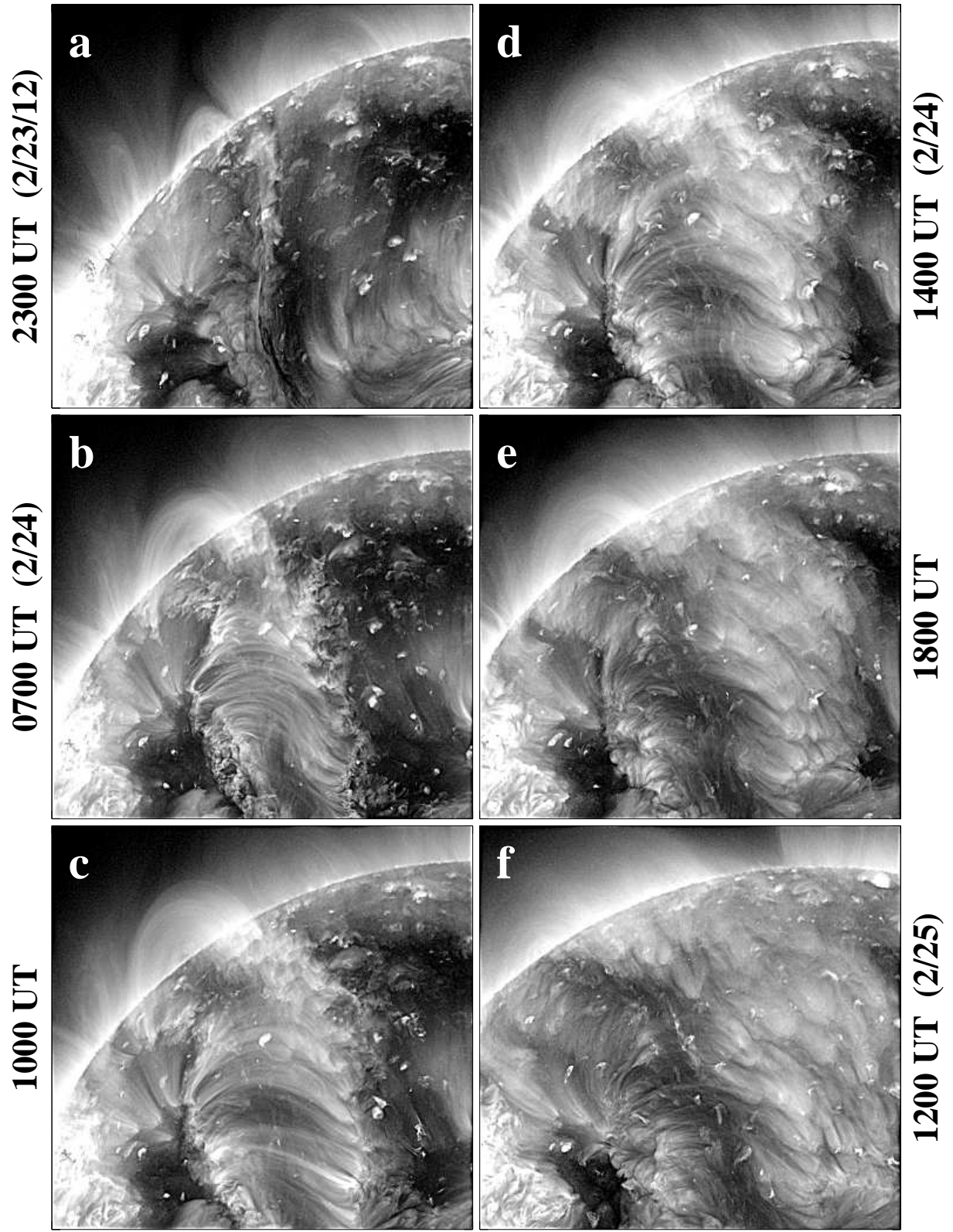}}
 \caption{193 \AA\ images, showing the restoration of cellular plumes after a
filament eruption. As times passes, initially skewed loops are replaced by higher,
less skewed loops, whose tops fade and whose legs remain as cellular plumes.}
\end{figure}

\end{document}